# Impact of Information Placement and User Representations in VR on Performance and Embodiment

Sofia Seinfeld, Tiare Feuchtner, Johannes Pinzek, Jörg Müller

**Abstract**— Human sensory processing is sensitive to the proximity of stimuli to the body. It is therefore plausible that these perceptual mechanisms also modulate the detectability of content in VR, depending on its location. We evaluate this in a user study and further explore the impact of the user's representation during interaction. We also analyze how embodiment and motor performance are influenced by these factors. In a dual-task paradigm, participants executed a motor task, either through virtual hands, virtual controllers, or a keyboard. Simultaneously, they detected visual stimuli appearing in different locations. We found that, while actively performing a motor task in the virtual environment, performance in detecting additional visual stimuli is higher when presented near the user's body. This effect is independent of how the user is represented and only occurs when the user is also engaged in a secondary task. We further found improved motor performance and increased embodiment when interacting through virtual tools and hands in VR, compared to interacting with a keyboard. This study contributes to better understanding the detectability of visual content in VR, depending on its location in the virtual environment, as well as the impact of different user representations on information processing, embodiment, and motor performance.

—————————— ◆ ——————————

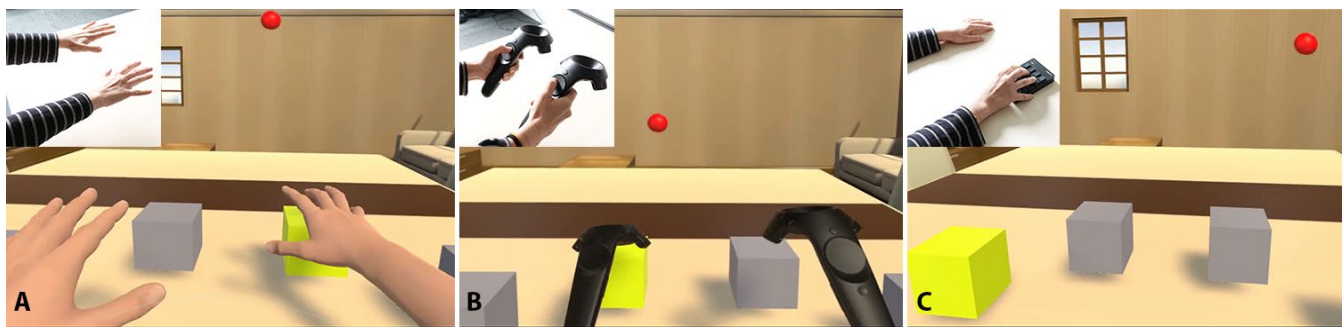

Figure 1. This study is based on a dual-task paradigm where participants have to quickly identify virtual targets (i.e., a glowing virtual cube) and simultaneously detect visual stimuli (i.e., a virtual red sphere appearing in different locations). Participants performed the task by touching the yellow cube with (A) Hands or (B) Controllers, or by pressing one of four respective keys on a (C) Keyboard. The red sphere that needed to be detected simultaneously can be seen floating in mid-air in each of the images.

## 1 INTRODUCTION

Designers of Virtual Reality (VR) interfaces need to consider where in a virtual environment information is presented for it to be noticed and processed efficiently. In particular, it may be desired to modulate the saliency of specific content or notifications. Some information is critical and should be noticed immediately, whereas other content should perhaps be placed in locations where it is less distracting. For instance, if users are executing a primary task in VR, they might still want to get notified about new emails, phone calls, calendar reminders, or events occurring in the real world, without it interfering with the task at hand.

In interaction with desktop computers or touchscreen devices, notifications are often provided visually and can vary in factors such as color, size, movement, and duration. In immersive VR notifications are no longer limited to a 2D screen surface but can be distributed throughout the virtual space. This allows designers to modulate notifications through additional factors, such as the user's spatial memory, distance perception, peripheral vision, etc. However, until now it is not well understood how the detectability of information in VR is affected by its location in virtual space and its distance with respect to the user.

Research has shown that cognitive factors, such as selective and divided attention, cognitive load, and

————————————————
- *Sofia Seinfeld is with the Institute of Computer Science, University of Bayreuth, Germany. E-mail: sofia.seinfeld@uni-bayreuth.de*
- *Tiare Feuchtner is with the Computer Science Department, Aarhus University, Denmark. E-mail: tiare.feuchtner@acm.org*
- *Johannes Pinzek is with the Institute of Computer Science, University of Bayreuth, Germany. E-mail: s2jopinz@stmail.uni-bayreuth.de*
- *Jörg Müller is with the Institute of Computer Science Department, University of Bayreuth, Germany. E-mail: joerg.mueller@uni-bayreuth.de*

perceived depth cues, influence the processing of visuo-spatial information [1], [2]. In visual search tasks, there is evidence that humans are faster at detecting targets located in the foreground compared to targets presented in the background, when provided with depth cues [3]. These studies have hypothesized that nearby visual information is processed more efficiently due to its ecological relevance, since it provides important information for immediate interaction with the environment. However, it is still not well understood whether these effects are transferable to immersive VR environments.

Additional aspects that might play a role in information detection and processing in VR are the way users are represented in the virtual environment (i.e., whether or not they directly control a virtual representation and the extent of their reach with it) and the type of input device used to control such a virtual representation. In the present paper, we describe the combination of these two aspects as "user representations" [4]. A series of studies has demonstrated that user representations in VR differently impact the degree of experienced embodiment [5], [6], multisensory integration [7], attentional processing [8], distance perception [9], size estimation [10]–[12], and task performance [13]. Hence, it is possible that the detection of visual content in an immersive 3D virtual environment is also differently modulated, depending on the type of user representation used to interact in VR. Moreover, the perceived mental workload while executing a task in VR and simultaneously detecting additional visual information may be influenced by the type of user representation, an issue which has not been researched yet. In this context, mental workload refers to the perceived cognitive, physical, and temporal demand of a task performed in VR [14].

Based on this previous research, the present study aims to contribute to a better understanding of the following three research questions:

- How is information in VR processed, when presented at different distances from the user and appearing in different areas of the user's field of view?
- Does the type of user representation (i.e., body, tools, or no visual representation) impact the detectability of visual stimuli in VR?
- What role do user representations (i.e., body, tools, or no visual representation) play, with respect to the subjective sense of embodiment and mental workload?

The subsequent section describes previous work in more detail, which has inspired and provides theoretical grounds for the present study, focusing on the processing of visuo-spatial information, user representations, and content placement in VR.

## 2 RELATED WORK
### 2.1 Detection of Visuo-Spatial Information
In everyday life we need to detect and process a vast amount of information embedded in the environment, with various stimuli competing for our attentional resources. In fact, divided and selective attention are critical abilities to effectively interact with the environment, enabling us to multitask and decide which information should be prioritized and acted on more quickly [15]. However, studies have shown that certain stimuli might be detected and processed more quickly, depending on their spatial location and how close they are to the user [3], as well as on other features, such as color, luminance, and motion [16].

Using a flicker paradigm with monocular and binocular cues, where participants are requested to detect color changes in different images, Ogawa & Macaluso [3] found that detection of changes in foreground/near objects is faster and more accurate compared to background/far objects. In accordance with these results, Qian & Zhang [2] also found higher detection sensitivity for changes in near depths compared to far depths in a change detection task including stereoscopic vision. Moreover, in 2D and 3D driving simulations, it has been shown that drivers are faster in detecting light changes at nearer depths compared to far away depths [1]. Such evidence points to a possible higher ecological relevance of nearby objects compared to faraway objects, since these may provide critical information and opportunities for immediate interaction with the environment.

To our knowledge, it has not yet been explored how visual stimuli are detected when located at different distances and visual angles from the user, while multi-tasking in VR. For instance, we are not aware of any study investigating how users detect visual stimuli in 3D space while simultaneously executing a primary motor task in immersive VR (e.g., quickly touching targets, while concurrently detecting visual stimuli appearing at different locations). The driving study by Andersen et al. [1] included a dual-task paradigm (i.e., detecting lights at different locations while driving in a simulator). However, this experiment was performed in a traditional desktop environment that was non-immersive. Furthermore, it has not been evaluated whether the awareness of sensory stimuli in different locations is also modulated by how the user is represented in VR. The visual appearance and input methods of a virtual representation have been shown to differently impact users' perceptions and behaviors [4], an aspect that we discuss in detail in the following section.

### 2.2 Virtual Representations of the User in VR
In VR experiences, the user is frequently represented through different virtual representations, which can be described as user representations [4]. User representations are the crucial element of the interface that allows users to carry out actions in the virtual environment and can drastically differ in terms of their visual appearance, the input method used to control them, and the mapping of controls. For example, users may be represented through virtual models of physical hand-held controllers. These virtual controllers typically look as if they were floating in mid-air (**Figure 1B**). However, interaction may also be enabled through direct keyboard or controller input, without users seeing any specific representation of themselves in the virtual environment (**Figure 1C**). An alternative way of representing the user is through realistic avatars that are experienced from a first-person perspective (**Figure 1A**). Here body tracking technologies allow direct control of the avatar's motions based on the user's own body movements. Such a representation typically leads to the embodiment of the virtual body: users experience that the virtual body is part of their physical body (i.e., body ownership), that they can control it (i.e., agency), and that they are collocated with the virtual body (i.e., self-location) [5], [6], [17].

Studies have demonstrated multiple advantages of interacting through a virtual body in VR. For example, Steed et al. [18] demonstrated that being represented by an active self-avatar, compared to a no-avatar condition, results in less mental load when doing a spatial rotation task and subsequently having to recall a sequence of letters. Mohler et al. [9] found that when participants are embodied in a fully tracked avatar, they are better at judging absolute egocentric distances, compared to when not having an avatar. Moreover, it has also been shown that having a self-avatar, or seeing an animated character in VR, improved performance in an interaction task within a virtual environment, compared to not having visual representation [19]. The results presented by Ebrahimi et al. [20] indicate that depth estimations are more accurate when interacting with high-fidelity avatars, compared to low-fidelity avatars or only seeing virtual end-effectors. Alzayat et al. [8] recently proposed a method to measure tool extension based on attentional processing. Their results suggest that users pay more attention to the task when using direct hand input, compared to controllers.

Using hand-held controllers as input device, Lougiakis et al. [13] found that visually representing the user with virtual hands in VR leads to stronger body ownership, in comparison to being represented by a virtual controllers or sphere. However, the sense of agency is equivalent for the different virtual representations. Moreover, in the same study it was found that, in a task where users had to select and move virtual objects, performance was better when interacting through virtual hands or controllers, compared to the sphere representation condition. Further, the controllers condition outperformed the hand and sphere representations in a positioning task. Ogawa et al. [21] found that the visibility and high anthropomorphism of self-avatars results in more realistic behaviors in VR (i.e., not penetrating a virtual wall), in contrast to when representing the user with virtual controllers or hands-only representations. Interestingly, no such differences were found by Lugrin et al. [22] during a VR game experience. In this study, body ownership, immersion, emotional and cognitive involvement, perceived difficulty, and game performance was not found to be significantly different for users interacting using virtual hands, virtual controllers, or a partially rendered virtual body. In all cases the virtual representations were controlled by means of hand-held controllers. When comparing interaction through direct hand gestures (i.e., using tracked gloves) to physical hand-held controllers, Lin et al. [23] showed that body ownership, the accuracy in judging the virtual hand size, and the perception of realism was stronger when using direct hand gestures. However, task performance was better when using physical controllers. With respect to keyboard input, it has been found that typing performance is significantly slower in VR when compared to real world text input, a limitation that might be overcome by the rendering of virtual hands and the keyboard [24]. However, it is not well understood how physical keyboard input compares to direct hand gestures or hand-held controllers.

Despite these varied findings presented in related work, we are not aware of any research addressing how being represented by an actively controlled self-avatar, compared to a virtual tool representation or not having any representation, impacts the processing of visual stimuli at different locations in a 3D space. Furthermore, it is not well understood how the type of input device used might impact information detection, task performance, workload, and embodiment, since several studies relied on a single input method, while only varying the visual representation of the user. Therefore, the present study aims to explore these phenomena in more detail.

## 2.3 Content Placement in VR

The question of where to place virtual content in VR has not yet been widely researched. Some recent studies have investigated the most effective ways of designing and delivering notifications in VR. For example, Ghosh et al. [25] explored the effectiveness of different designs for audio, tactile, and visual notifications in VR, and they highlight that it is critical to consider their position with respect to the user's body. Their findings indicate that notifications are effectively detected when collocated with the virtual controllers, however, with the limitation that controllers are often outside the user's field of view and are used to execute other actions. Moreover, when notifications were presented very close to the user's body, they occasionally caused jump scares. This effect might be related to the brain's bodily and spatial representation mechanisms that enable humans to quickly react when a stimulus is registered near the body, to protect it from external threats [26].

Rzayev et al., [27] further investigated this topic, focusing specifically on the placement of notifications in different virtual environments, while users executed various types of tasks. The authors placed notifications either 25 cm in front of the headset (heads-up display), 15 cm away from the virtual representation of controllers (on-body), on the nearest wall in the virtual environment (in-situ), or floating in front of the user at varying distances. Their results indicate that notifications in the heads-up display (i.e., 25 cm in front of the headset) were more quickly detected but were also perceived as more intrusive. In-situ notifications (i.e., on the nearest wall in the virtual environment) resulted in longer reaction times and more misses. No clear differences in reaction times or hits were observed when notifications were floating in front of the user or when they were placed 15 cm away from the virtual representation of controllers (i.e., on-body).

Overall, studies researching notifications in VR highlight the importance of effectively selecting the placement of virtual content with respect to the user's body. However, we are not aware of any studies that systematically explore the effect of the distance from the user at which notifications are presented. Moreover, notifications consisting of diverse types of content, e.g., ranging from email notifications to a virtual character that talked to the user [27], may introduce confounding variables. Finally, there is no evaluation of how different user representations might impact the detection of notifications [4]. The present study aims to better understand content placement in VR, by controlling for the above-mentioned factors.

## 3 EXPLORING INFORMATION PLACEMENT AND USER REPRESENTATIONS

### 3.1 Experimental Design and Hypothesis

We designed a VR experience based on a dual-task paradigm: Users were instructed to quickly execute a motor task (Cube Task), while simultaneously detecting visual information presented in different locations of the 3D space (Sphere Task). Descriptions of the Cube and Sphere Tasks are given in section 3.3 (Experimental Interaction Tasks).

The present study involves a fully counterbalanced within-groups experimental design, featuring the dual-task paradigm in three experimental conditions. In these conditions participants executed the Cube Task either through i) a virtual body (i.e., Hands), ii) virtual tools (i.e., Controllers) or iii) Keyboard input (no visual representation). The user representation in each experimental condition is shown in **Figure 1**.

Based on experimental evidence from related work, the present study aims to test the following hypotheses:

**H1:** When information is presented near to the user's body, it is more quickly detected (higher detection performance), compared to when information is presented further away in the virtual environment.

**H2:** Representing the user by means of a virtual body leads to an enhanced sense of embodiment and decreased mental workload, compared to the use of a virtual tool (i.e., controllers) or a physical keyboard. In relation to motor performance when interacting through the different user representations we did not have any particular preestablished hypothesis.

### 3.2 Implementation and Interaction Techniques

The VR scene and experimental setup was implemented using *Unity3D[1] (version 2019.1.5f1)*. Users experienced the virtual environment through an *HTC Vive[2]* Head Mounted Display (HMD).

Different additional equipment and mappings were used, depending on the type of user representation in the experimental conditions:

- *Hands:* In this condition, participants saw a pair of virtual hands that were collocated with respect to their real hands. Users' hand and finger movements were tracked in real time using a *Leap Motion[3]* sensor that was attached to the front of the HMD. This provided participants with visuo-motor feedback, by linking their real hand movements to those of the virtual hands. During the Cube Task (details below), users interacted with the virtual environment by touching targets with these virtual hands (**Figure 1A**).
- *Controllers:* Participants saw a pair of virtual controllers floating in mid-air, which were collocated with physical hand-held HTC Vive Controllers. In the Cube Task, users touched virtual targets with these virtual controllers (**Figure 1B**).
- *Keyboard:* In this condition, no visual representation of the user was provided in the virtual scene. During the experiment users selected virtual targets in the Cube Task by pressing four corresponding keys on the keyboard (**Figure 1C**).

The two conditions with virtual user representations feature the virtual HTC Vive controller model included in the SteamVR plugin for Unity (v1.2.2), as well as an adaptation of a virtual hand model from the Leap Motion Hands Module (v2.1.3). In both conditions, touching of targets was detected using Unity's built-in collider system. For this purpose, BoxColliders were attached to the virtual hands and the virtual controllers. To achieve a comparable level of difficulty in the Cube Task between these conditions, the attached colliders were of equal size. When one of these colliders intersected with the BoxCollider of a target cube, the respective cube was selected. In the Keyboard condition, participants had the fingers of their dominant hand resting on a number pad, on which the second farthest row of keys (from left to right: '7', '8', '9', '-') was mapped to the four cubes correspondingly, so that pressing '7' selected the left-most, '-' selected the right-most cube, etc. The latencies of the different input methods were measured following the procedure described in Müller et al [28], taking 20 measurements per input method. The differences in latencies for each input method were small, in the range of milliseconds, and therefore not perceptible to the user (Hands latency=75.30ms, std: 6.76; Controller latency=36.05ms, std: 3.89; and Keyboard latency=79.25ms (std: 6.47).

The virtual environment was designed to resemble a living room, in which the participant appeared to be sitting at a table.

### 3.3 Experimental Interaction Tasks

The user study involved a dual-task paradigm, i.e., the user was required to execute two tasks simultaneously, namely the Cube Task and the Sphere Task. Both tasks are described in detail below.

#### 3.3.1. Cube Task

We designed a motor task based on selecting virtual targets, by touching them with the virtual a) Hands or b) Controllers or selecting them through c) Keyboard input. In the virtual scene, participants saw four targets in the form of grey cubes, which were placed on a table in front of them within arm's reach (**Figure 1**). In random order, one of the cubes turned yellow and participants were instructed to select the highlighted cube as quickly as possible. As soon as the yellow cube was selected, its color changed back to grey and another cube in a different location immediately turned yellow instead. The four cubes were highlighted in random order, but the same cube could not light up twice in a row.

---

[1] https://unity.com/
[2] https://www.vive.com/us/product/vive-virtual-reality-system/
[3] https://www.leapmotion.com/

### 3.3.2. Sphere Task

For the Sphere Task, we requested participants to pay attention to visual stimuli (i.e., the appearance of a red sphere) and press a pedal with their right foot as soon as they detected such a stimulus anywhere in the scene.

We defined 32 fixed sphere positions in the virtual environment, at four different distances from the user: 15 cm, 60 cm, 285 cm, 485 cm. At the first two distances the sphere is within easy reach of the user's arms, while at the latter two it lies far beyond. At each of these four distances, sphere positions were distributed across four horizontal locations (horizontal angles) and two possible elevations (vertical angles), based on a 3D spherical coordinate system (see **Figure 2**). Horizontally, sphere positions were either located in the user's central field of view (hor. angle: 80° and 100°), or peripheral field of view (hor. angle: 60° and 120°). Vertically, 50% of targets were arranged at the user's eye level (vert. angle: 90°), where a stimulus may be assumed to be most noticeable. The remaining targets were positioned at a higher elevation (vert. angle: 80°) to explore noticeability of stimuli further towards the user's peripheral vision. No sphere positions at lower levels were tested (e.g., at same elevation as cubes or lower), since here a stimulus was prone to occlusion by the cubes or the virtual desk. The specification of these sphere positions was informed by pilot studies, aiming at distributing stimuli throughout the user's field of view, such that they were all simultaneously visible when looking straight ahead. Positions were fixed in world space (i.e., did not move when the user moved the head), in relation to the user's assumed forward gaze direction when interacting with the cubes. During the experiment, participants were instructed to avoid strong head rotations in order to not get distracted from the tasks at hand (i.e., Cube and Sphere tasks).

Since we wanted to assess the processing of visual information based on its proximity to the user's body, but not based on varying size or salience of the stimuli, we controlled for the potential influence of such additional factors. We kept the size of the sphere constant in terms of the visual angle, irrespective of its distance to the user. The sphere was rescaled to cover the same visual angle of 9.46° for the different distances. Therefore, the sphere was smaller at closer distances and bigger at further distances in order to cover the same visual angle. Subjectively, this made distant spheres seem larger than close ones, due to our interpretation of depth cues that objects become smaller as they move further away. Importantly, we found no indication that participants' depth perception was hampered by the increased size of far-away objects. Depth perception was supported through binocular disparity and motion parallax from minor head movements during the task, as well as shading and specular highlights. No difficulties were reported in distinguishing the various

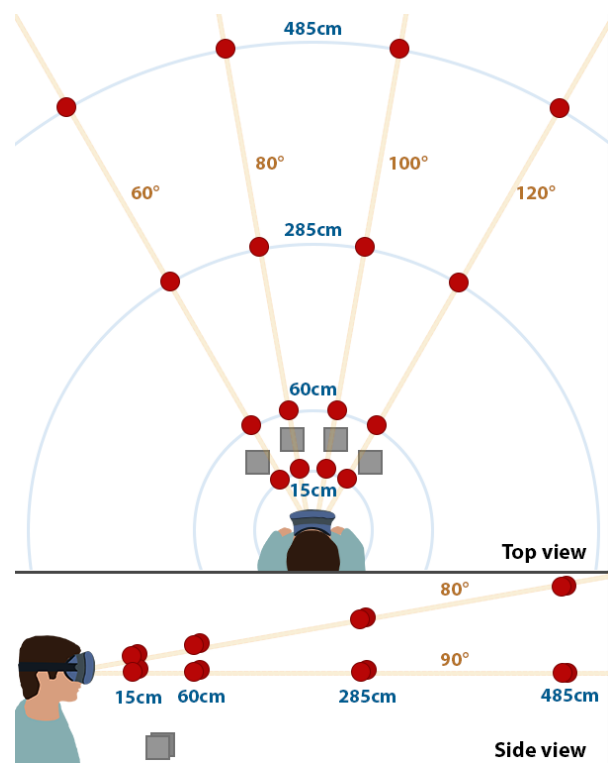

Figure 2. Sphere positions were defined at four different distances from the user (15 cm, 60 cm, 285 cm, 485 cm). These were distributed at four horizontal angles (central: 80°, 100°; peripheral: 60°, 120°) and 2 vertical angles (80°, 90°), resulting in 32 possible positions. The four targets of the Cube Task were positioned on the desk, within arm's reach of the user (illustrated as gray cubes).

distances at which the sphere appeared during the pilot and experimental studies.

During the task, a single sphere repeatedly appeared at a randomly chosen predefined position and was presented for a maximum duration of 2500 milliseconds. Participants were requested to press the foot pedal with their right foot as soon as they detected the red sphere, upon which the sphere would disappear. If the sphere remained undetected, i.e., the presentation duration elapsed without the participant pressing the pedal, the sphere disappeared automatically. After a short interval, the sphere then reappeared at a different location. This interstimulus interval ranged from 3000 to 5000 milliseconds and was randomly varied. The specific timings, duration of each stimulus appearance, and interstimulus intervals were chosen based on the detection response task paradigm [29]. This is a validated paradigm to assess the impact of cognitive load when executing two simultaneous tasks (e.g., driving a car while perceiving different sensory stimuli).

### 3.4 Participants

A total of 24 participants (Mean Age=24.46, Standard Deviation=3.62, 12 Male, 1 left-handed) took part in the study. Inclusion criteria for participants involved not suffering from any type of sensory impairment, no neurological disease, and no intake of psychoactive medications. This study was granted ethical approval by the Ethics Committee of the University of Bayreuth and followed ethical standards according to the Helsinki Declaration. Participants received financial compensation for their participation.

### 3.5 Procedure

After signing their informed consent, participants were immersed in the virtual environment. The experiment then started with a baseline condition, in which baseline detection performance measures were recorded for the Sphere Task: participants were asked to detect the virtual sphere appearing at different locations in the 3D space in a single-task paradigm, i.e., without simultaneously executing the Cube Task. Participants were tasked to detect the sphere with highest possible speed and accuracy. The sphere was presented twice in each of the 32 unique locations in random order, resulting in a total of 64 trials. No virtual user representation was given during this task, and the only required input was provided with the foot pedal.

The same detection response measures were again recorded in the experimental conditions with the dual-task paradigm when users simultaneously executed the Sphere and Cube Task in three fully counterbalanced experimental conditions. The visual stimulus again appeared twice at all 32 positions in random order. This results in 64 trials per condition with a factor design of 3 (user representations) x 4 (distances) x 2 (vertical angles: high/low) x 2 (horizontal angles: 2 central/2 peripheral). In each of these trials, the appearance of the sphere (i.e., Sphere Task) was independent from a different cube lighting up (i.e., Cube Task). Participants were asked to detect spheres with highest possible speed and accuracy. However, they were also told to touch the highlighted cube as quickly as possible. Since a new cube lit up immediately after the old one was selected, this motor task demanded a higher pace compared to the detection of the sphere, which appeared at longer and unpredictable intervals. Consequently, the Cube Task competed with the Sphere Task for the user's attention. The Cube Task was repeated, until all trials of the Sphere Task were presented (i.e., sphere appearing at different locations).

Immediately after completing each experimental condition, participants completed a digital version of the VR questionnaire and NASA TLX questionnaires, as described below. The study lasted approximately one hour.

### 3.6 Measurements

#### 3.6.1 Detection Performance

In the Sphere Task, detection performance was evaluated based on the detection rate and response time. Detection rate was measured by counting the number of times the sphere was correctly detected within the defined time frame (i.e., 100ms-2500ms after appearing). Response time was measured as the elapsed time between the sphere appearing and the participant pressing the foot pedal.

#### 3.6.2 Motor Performance

We evaluated motor performance through the Cube Task. From this task, we recorded the number of hits defined by the number of highlighted virtual cubes (i.e., targets) touched during each experimental condition. A hit was only counted if the correct cube was touched. Errors were not counted.

TABLE 1. QUESTIONNAIRE ITEMS INCLUDED IN THE VR QUESTIONNAIRE.

| Variable | Questionnaire Item |
|---|---|
| *body ownership* | I had the illusion that the *virtual hands / virtual controllers / keyboard* were my own hands. |
| *agency* | I felt as if the movements of the *virtual hands / virtual controllers / keyboard* were caused by my movements. |
| *control* | I felt like I could control the *virtual hands / virtual controllers / keyboard* as if they were my own hands. |
| *self-location* | I felt as if my hands were located where I saw the *virtual hands / virtual controllers / (where I felt the) keyboard* |
| *effectiveness* | I felt very effective selecting the cubes with the *virtual hands / virtual controllers / keyboard.* |
| *realism* | The virtual experience felt real. |

#### 3.6.3 VR Questionnaire

In a questionnaire we included a series of items relating to different aspects of the VR experience (see **Table 1**). These questions were rated on a 7-point Likert scale, with 1 signifying complete disagreement and 7 complete agreement with the statement. The included question on body ownership relates to the perceptual illusion of feeling that the virtual object was part of the real body [30], [31]. The agency and control questionnaire items aimed to assess participants' sense of being responsible for effecting changes in the virtual environment and being able to control their user representations. We also included a specific question to measure whether participants felt that their own body was located where they perceived their user representation to be, namely self-location. Finally, the questions on effectiveness and realism refer to the extent to which participants felt that they could effectively execute the tasks and the degree of immersion they experienced, respectively. All of these questions have been used previously in similar research [17,32].

#### 3.6.4 NASA Task Load Index (TLX)

Participants also completed the NASA TLX [33] which is a validated questionnaire for evaluating mental workload. It assesses six sub-scales: mental demand, physical demand, temporal demand, perceived performance, effort, and frustration. This index has been widely used in human-computer interaction to measure workload during interaction [14], and in this study we wanted to explore whether subjective workload could vary depending on the type of virtual user representation used in a dual-task paradigm.

## 3.7 Statistical Analysis

Generalized Linear Mixed Models (GLMM) were used to analyze the time taken to detect the red sphere during the baseline and the different experimental conditions. In accordance with recommendations by Baayen & Milin [34] and Lo & Andrews [35], we used a gamma function to fit the distribution of the model to the reaction time data, since reaction times do not follow a normal distribution. To analyze the impact of the different input methods and virtual representations of the user, we first ran a GLMM including the factors User Representation (hands, controllers, and keyboard), Distance (15 cm, 60 cm, 285 cm, and 485 cm), Vertical Angles (80° and 90°) and Horizontal Angles (central: 80°, 100°; peripheral: 60°, 120°) as the independent variables of the model. The participants' IDs were set as random effects. For the baseline trial and also for the actual experimental trials, we ran an additional GLMM, where Distance, Vertical Angles, and Horizontal Angles were set as the independent variables and participants' IDs were defined as random effects in the model, but the factor User Representation was not included. This additional analysis allowed us to evaluate the performance in detecting visual stimuli (i.e., spheres) based on their location within the virtual environment, independently of how the user was virtually represented and of the input method used. All statistical analyses and plots were performed using Stata v.16.1.

## 4 RESULTS

The following results regarding detection and motor performance stem from the analysis of only 23 participants (n=23), since one participant was excluded from the analysis, due to an error in the data collection. The analysis of questionnaire responses includes all participants (n=24).

### 4.1 Detection Performance in Sphere Task

#### 4.1.1 Visual Information Processing Baseline

In the baseline trial, we found no significant main effect or interactions of Distance, Vertical Angles, or Horizontal Angles on the time taken to detect visual stimuli (i.e., red spheres) at different locations. The detailed results of the GLMM model can be seen in Table S3 of the Supplementary Materials. **Figure 3** shows average times to respond to the sphere appearing at different distances. The detection rate in the baseline task was 100% (i.e., the stimulus was always successfully detected).

#### 4.1.2 Visual Information Processing during Cube Task

In the full GLMM model including the factors User Representations, Distance, Vertical Angles, and Horizontal Angles, no significant main effect or interaction effects were found for the time taken to respond to visual stimuli in a dual-task paradigm. However, we did observe a trend indicating lower response times at a distance of 15 cm compared to 285 cm (p=0.068) and 485 cm (p=0.081), as well as between a distance of 60 cm compared to 285 cm (p=0.072) and 485 cm (p=0.086). No significant difference in response times were observed between 15 cm and 60 cm (p=0.977) or between 285 cm and 485 cm (p=0.934). See detailed results of the GLMM in Table S2 of Supplementary Materials.

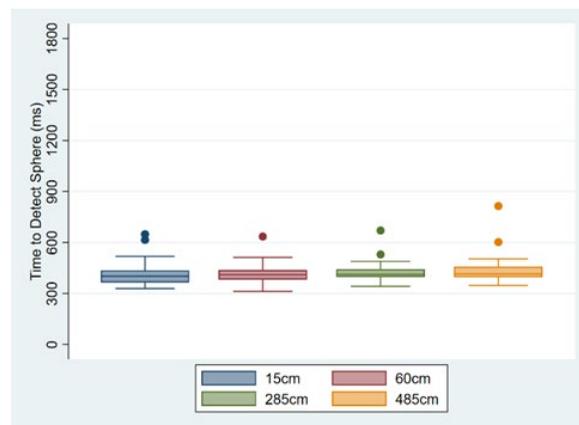

Figure 3. Boxplot showing the average detection time for visual stimuli by distance during the baseline trial.

To better understand the impact of Distance, Vertical Angles, and Horizontal Angles on the detectability of visual stimuli in VR, we ran an additional analysis excluding the factor User Representations. We decided to exclude the factor User Representation from this analysis since it did not show any significant main effects or interactions in our previous model. We found a significant main effect of Distance, indicating that the time to respond to visual stimuli was lower when the red spheres were located at distance of 15 cm compared to far distances of 285 cm (p<0.01) and 485 cm (p<0.01). Response times were also significantly lower for stimuli placed at 60 cm, when compared to a 285 cm (p=0.032) and 485 cm (p<0.01). No significant differences were found between stimuli located at 15 cm and 60 cm (p=0.182) or between stimuli placed at 285 cm and 485 cm (p=0.469). Furthermore, no significant main effects or interactions were found for the factors Vertical Angles and Horizontal Angles. **Figure 4** shows the time taken to respond to visual stimuli (i.e., spheres) at different distances across all conditions and **Table 2** shows the detailed results obtained through the GLMM for the factor Distance. The complete results of the GLMM can be found in Table S3 of Supplementary Materials.

TABLE 2. RESULTS OF THE GLMM FOR THE FACTOR DISTANCE ACROSS ALL CONDITIONS.

| Distance 1 | Distance 2 | z | P>|z| | [95% Conf. Interval] |
|---|---|---|---|---|
| *15 cm* | *60 cm* | 1.34 | 0.182 | -0.0181917 - 0.0959535 |
| | *285 cm* | 3.49 | 0.000 | 0.0444192 - 0.1585645 |
| | *485 cm* | 4.21 | 0.000 | 0.0654883 - 0.1796335 |
| *60 cm* | *285 cm* | 2.15 | 0.032 | 0.0055384 - 0.1196836 |
| | *485 cm* | 2.87 | 0.004 | 0.0266074 - 0.1407527 |
| *285 cm* | *485 cm* | 0.72 | 0.469 | -0.0360035 - 0.0781417 |

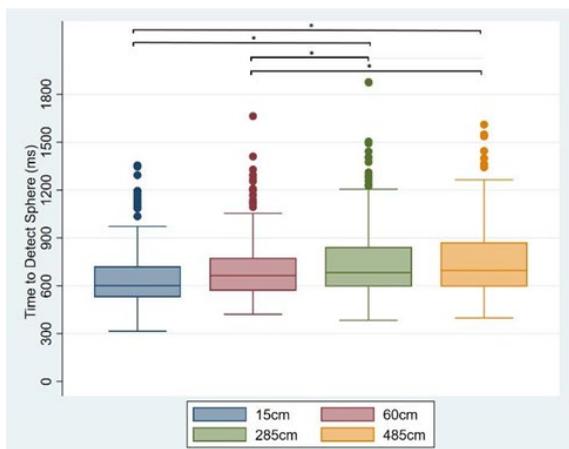

Figure 4. Boxplot showing the response time to visual stimuli by distance from the user, with data aggregated for all three conditions (i.e., all user representations).

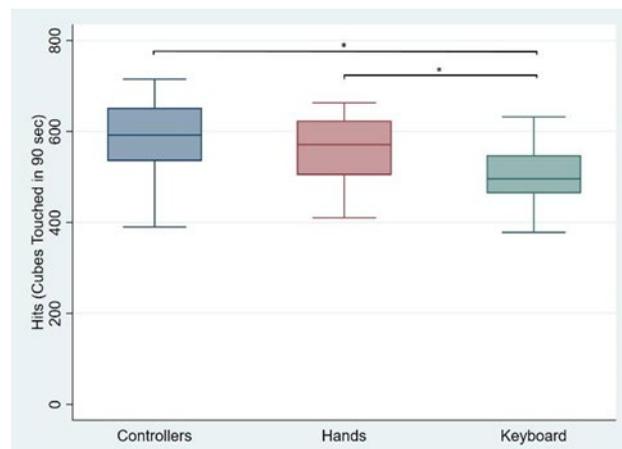

Figure 5. Average number of hits (cubes touched) in each experimental condition (User Representation).

Overall, the number of undetected stimuli was very low. Per distance and across all conditions, the sphere appeared 1104 times, and averaging across all conditions participants only missed 2.08% stimuli at 15 cm distance, 1.81% at 60 cm, 5.43% at 285 cm, and 5.34% at 485 cm. This means that in most trials participants were successful in detecting the sphere every time it appeared.

The total time taken to complete the task was approximately 5 minutes for all the experimental conditions (Hands duration= 311sec, Controllers duration= 309sec, Keyboard duration= 307sec).

### 4.2 Motor Performance in the Cube Task

Analyzing participants' performance in the Cube Task, a repeated measures ANOVA shows a main effect of Condition (Hands, Controllers, or Keyboard) on the number of hits (i.e., highlighted virtual cubes touched) ($F(2,44)=15.01$, $p<0.01$, partial $\eta_{per2}=0.41$). Further post-hoc comparisons with Bonferroni corrections showed that participants accomplished a significantly higher number of hits with virtual Hands ($p<0.01$) and Controllers ($p<0.01$), compared to the Keyboard. No significant difference in number of hits was found between the Hands and Controllers (see **Figure 5**). The residual errors of the ANOVA were normally distributed.

### 4.3 VR Questionnaire

#### 4.3.1 Body Ownership, Agency, and Self-Location

Friedman tests indicate that there was a significant difference in feeling of body ownership ($\chi^2=30.50$, df=2, $p<0.001$) and agency ($\chi^2=11.56$, df=2, $p<0.01$) between conditions (i.e., user representations). As reflected in **Figure 6**, further Wilcoxon signed-rank tests reveal that reported body ownership ($p<0.01$) and agency ($p<0.01$) scores were significantly higher when interacting using virtual Hands compared to a Keyboard. Participants also reported higher body ownership for the virtual Hands compared to the Controllers ($p<0.01$), whereas these conditions did not differ in terms of perceived sense of agency ($p=1.00$). No significant differences in body ownership ($p=0.08$) and agency ($p=0.09$) were found between the Controllers and the Keyboard conditions. No significant differences were found between user representations in self-location (Friedman Test; $\chi^2=3.65$, df=2, $p=0.16$).

#### 4.3.2 Control, Realism, and Effectiveness

In terms of perceived control, there was a clear difference between conditions (Friedman Test; $\chi^2=20.73$, df=2, $p<0.001$), which can also be seen in **Figure 7**. Wilcoxon signed-rank tests indicate that the degree of perceived Control was comparable when interacting with Hands and Controllers ($p=0.13$). However, participants reported a higher sense of control when using the Hands ($p<0.01$) and Controllers ($p<0.01$), compared to the Keyboard. Participants also perceived significant differences in terms of realism ($\chi^2=7.49$, df=2, $p=0.02$) and effectiveness ($\chi^2=9.74$, df=2, $p<0.01$) of the interaction. Wilcoxon signed-rank tests indicate that the Hands condition was perceived as more realistic when compared to the Keyboard ($p<0.01$), with no differences between the Hands and the Controllers ($p=0.59$), or between the Controllers and Keyboard ($p=0.13$). For perceived effectiveness we found that ratings were higher when participants interacted through the virtual Hands ($p=0.02$) or the Controllers ($p<0.01$), compared to Keyboard input. Again, there were no differences in reported effectiveness when comparing the Hands and Controllers (p=0.55).

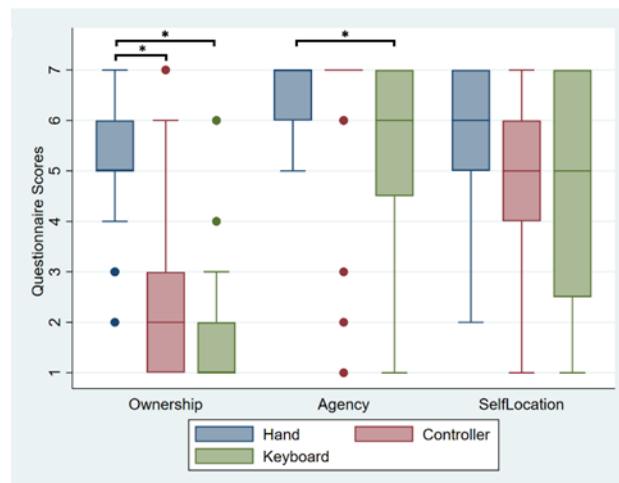

Figure 6. Boxplot with Likert scale ratings for Body Ownership, Agency, and Self-location items in each of the user representation conditions.

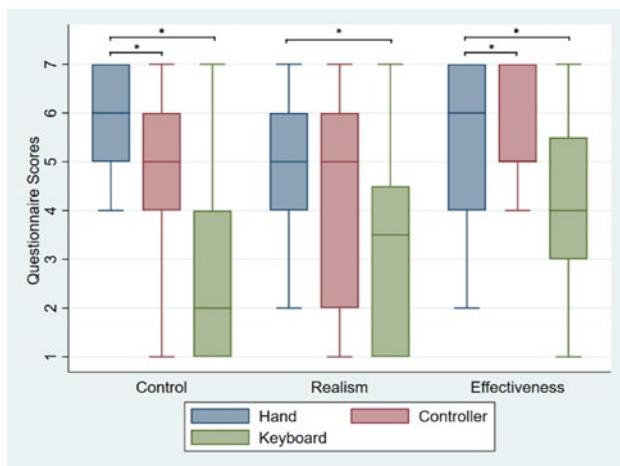

Figure 7. Boxplot with Likert scale ratings for Control, Realism and Effectiveness items in each of the conditions.

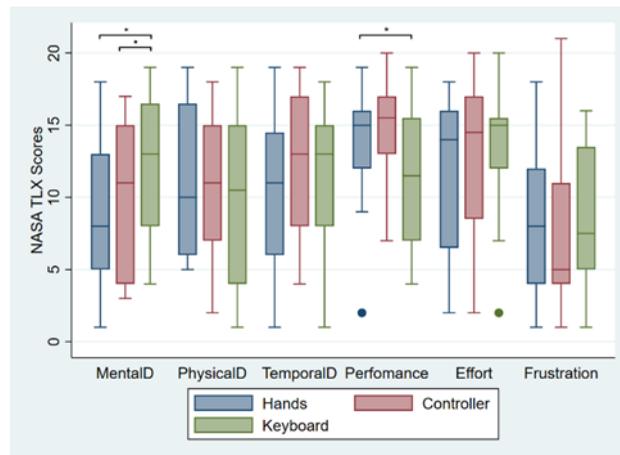

Figure 8. Boxplot with results from the NASA TLX questionnaire, showing differences between the conditions.

### 4.4 NASA TLX

According to Shapiro-Wilk tests, the data of the different NASA TLX sub-scales was not normally distributed. For this reason, this data was analyzed using Friedman tests. When a significant difference was found, we further explored the differences between conditions using pairwise comparisons with Wilcoxon signed-rank tests. **Figure 8** summarizes the scores given by participants in the different NASA TLX subscales, based on the type of user representation (i.e., condition).

We found a significant difference between conditions in Mental Demand (Friedman: $\chi^2$=11.93, df=2, p<0.01) and Performance (Friedman: $\chi^2$=11.93, df=2, p<0.01). Pairwise comparisons reveal that Mental Demand was significantly higher in the Keyboard condition, when compared to both the Hands (p<0.01) and the Controllers (p=0.04) conditions. Further, subjective Performance was higher for the Hands (p<0.01) and Controllers (p<0.01) conditions, compared to the Keyboard. No significant difference between the Hands and Controllers were found in these measures.

We did not find any further significant differences in any of the remaining NASA TLX sub-scales: Physical Demand (Friedman: $\chi^2$=3.05, df=2, p=0.22), Temporal Demand ($\chi^2$=1.27, df=2, p=0.53), Effort (Friedman: $\chi^2$=2.92, df=2, p=0.23), and Frustration (Friedman: $\chi^2$=2.63, df=2, p=0.27).

## 5 DISCUSSION

The evidence from our study strongly supports our first hypothesis (H1), since detection performance of visual information was significantly influenced by distance, with information presented near the user's physical body being more quickly detected. However, this effect occurred only in the dual-task paradigm, i.e., when users had to quickly touch glowing targets while concurrently detecting visual stimuli appearing at different locations, requiring selective and divided attentional mechanisms. No such results were observed in the baseline control condition, where only selective attention was required for detecting visual stimuli appearing at different locations. It is plausible that having more cognitive resources available during the baseline condition, since no concurrent motor task had to be performed, resulted in a more efficient processing of visual stimuli appearing at different distances. Conversely, the limited number of cognitive resources in the dual-task paradigm may have led to a prioritization of closer stimuli [15]. Moreover, we did not find evidence supporting the notion that detectability of visual information in a 3D immersive virtual environment is differently modulated by the type of user representation (i.e., virtual representation of the user and the input device used to control the representation). In relation to our second hypothesis (H2), we found that users indeed reported a strong sense of embodiment and lower mental workload when interacting with virtual Hands, but to some extent this was also the case when using Controllers. Finally, we also found that interacting through virtual Controllers positively impacted motor performance, analyzed by the number of hits, compared to using a Keyboard. However, no such difference in motor performance was found between Controllers and Hands.

Our findings support the notion that when faced with more than one task in VR, the processing of nearby virtual objects is prioritized and these are detected more quickly, compared to virtual objects located further away. Such evidence is in line with past studies showing that visual changes in an image are noticed more quickly when they occur in the foreground compared to the background [3], or that near-depth changes in a stereoscopic image are noticed faster than when occurring at far depth [2]. Moreover, these results are also in accordance with a study that found drivers to be faster at detecting light changes at nearer depths compared to far away depths while driving in a non-immersive 3D environment [1]. However, in the present research we expand these results to an immersive VR scenario. Overall, this evidence highlights the ecological relevance of nearby objects in real and virtual environments since they enable immediate interaction with the surroundings.

The enhanced processing of nearby visual information might also be related to Peripersonal Space (PPS). The PPS is the space immediately surrounding a person's body, which plays an important role in linking the user's body position to different sensory stimuli located in the environment. Interestingly, it has been shown that PPS boundaries in immersive VR are consistent with those of the real world and can be impacted by interactions with agents and objects [36,37]. Moreover, several studies have shown enhanced multisensory integration when information is presented within the PPS [37–39]. The increased multisensory processing within the PPS has been attributed to the role this area plays as a protective buffer zone, preventing harm to the body [26]. Moreover, studies have also shown that the PPS is the interface between perception and the actions executed in the environment through the body or through a manipulated physical or virtual tool [7, 40,41].

However, based on the experimental design used in the present study we cannot clearly elucidate whether the PPS played some role in the present findings. Future research should include tasks where more than one sensory modality is stimulated (e.g., crossmodal congruency task) and also manipulate the distance in which the primary task is performed (e.g., performing the task of touching the cubes at different distance such as 15 cm, 60 cm, 285 cm or 485 cm) to assess whether the PPS played some role in the present findings and in the processing of the visual stimuli presented at different distances.

Also in accordance with our results, the studies of Ghosh et al. [25] and Rzayev et al. [27] show that in VR notifications are more quickly detected when presented near to the user's body (e.g., on controllers or in front of the face). However, in these studies, stimuli presented very near to the participants were also perceived as more disruptive. It is therefore recommendable to present virtual content at a comfortable distance from the users unless information of great importance must be communicated. It should be noted that, although in our study users took longer to detect visual information that was located further away, the sphere was still correctly detected in the vast majority of cases (i.e., very few misses). This suggests that in a simple VR scenario, placing virtual content in the background may still result in effective, albeit slower, detection. However, this may not be the case in more complex VR scenes that include further distractors and complex interactions, which might hinder the detectability of content presented in the background. This should be considered when designing immersive virtual scenarios, in order to decide on the location of virtual notifications and virtual objects. However, further research is needed to understand other factors that might impact the detectability of visual stimuli in VR. For instance, Hillaire et al. [16] introduced a visual attention model of visual exploration in 3D environments, which includes bottom-up (e.g., colors, luminance, depth, and motion) and top-down factors (e.g., type of task). We believe that the results of the present research could help further improve the performance of such models in creating effective saliency maps.

Moreover, it is possible that detectability of visual stimuli within a VR scene is also modulated by the type of virtual environment rendered. Creem-Regehr et al. [42] demonstrated that users´ egocentric distance perception is differently impacted when immersed in an indoor or outdoor environment, with larger underestimations of distance occurring in outdoor virtual environments. Therefore, the detectability of visual stimuli within a VR scene might similarly be impacted by the type of virtual environment represented (indoor or outdoor) in scenarios where distance does indeed modulate visual perception (e.g., dual-task paradigm). Furthermore, in the present study no significant differences were found in the detectability of visual stimuli presented in central vs. peripheral vision, despite past evidence showing that virtual objects located on the sides of a VR environment are perceived as being further away compared to central objects [43]. This suggests the possibility that the anisotropy effect found in VR when perceiving distances is not strong enough to impact visual stimulus detectability. Future studies should aim to test these theories to better understand the different aspects that modulate visual stimulus detectability in VR.

Importantly, the present research tries to control for several factors that may influence the processing of visual information in VR. First, a single type of simple visual stimulus was used (i.e., red sphere). This allowed us to test visual processing of stimuli based only on distance and the user representation, without potentially confounding effects from cognitive processing that may be involved when presenting complex stimuli (e.g., email notifications or a talking avatar). Second, we systematically controlled for the size of the visual stimuli, to ensure that they covered the same visual angle at all distances. Third, the presentation duration of stimuli was based on the detection response task [29], a well validated measure for assessing mental workload when multitasking. Finally, for the first time, we have researched how visual stimuli are processed at different distances in connection to the virtual representation of the user. Nevertheless, despite the care taken to control several potential confounding effects there are still factors that could have played a role in the observed results. For example, when experiencing VR through an HMD, we are prone to suffer from the Vergence-Accommodation (VA) conflict [44]. The focus distance of virtual objects, which the eyes have to accommodate to, is usually fixed at infinity. In contrast, the eyes' vergence is adjusted differently for close objects, such that a conflict between vergence and accommodation arises. This leads to contradicting depth cues. However, we observe a completely different pattern of results between the baseline (i.e., single-task) condition and the experimental trials (i.e., dual-task paradigm), in particular regarding the effect of distance on detectability of stimuli. Thus, we deem it unlikely that the VA conflict accounts for the present findings, since we would expect the same patterns in the baseline condition, as well as in the experimental trials based on the provided depth cues being identical.

In accordance with past research, and supporting our second hypothesis, we have found that the type of user representation appears to modulate embodiment [4], [6].

Moreover, the present findings indicate that direct manipulation through Hands and Controllers leads to better motor performance in the Cube Task, compared to interaction through Keyboard input. It may be argued that the lower performance with the Keyboard could be due to the user having to learn a more complicated mapping (assigning keys to cube locations), which may require further cognitive and motor processing. This is also supported by the increased mental workload found in this condition. If this were the case, the results could be expected to improve with training, which should be further explored in future studies. A further aspect which might have impacted the results is the fact that in the keyboard condition users only used one hand to press keys, while in the Hands and Controllers conditions participants used both hands to hit the cubes. Using a different type of input device (i.e., tablet and stylus) it has been shown that a bimanual technique leads to better performance than an unimanual input modality, under increased cognitive load [45]. This might also be the case in the present study, therefore future research should address whether performance in VR based on keyboard inputs improves when the user is allowed to use both hands in comparison to only one.

Finally, we would like to highlight that reaching for targets in VR using our hands or hand-held tools parallels the actions we carry out in the physical world, making it well suitable for novice users. Moreover, virtual hand representations have also been shown to result in perceptual improvements in previous research [46], [47]. Thus, interactions are potentially more intuitive [48], require lower cognitive load, and lead to higher presence in VR [21], when direct manipulation is involved. Finally, in accordance with other studies, we have found that the sense of body ownership is strongest when interacting through a virtual body (i.e., Hands) [49]. To some extent, users also experienced a sense of embodiment over the Controllers, in perceiving high agency and control over the actions executed through them, as well as self-location at the controllers' position.

## 6 FUTURE WORK

The user study presented in this paper explores information presentation in VR from the perspective of visual processing, depending on stimulus distance and the impact of user representations. There are several further aspects that future research should explore. For example, processing of different modalities (audio/visual), or processing visual stimuli of higher complexity (e.g., email notifications or animated objects) might be different. Further, it remains to be studied whether the difference in processing of near and far stimuli depends on the primary task being a motor task, or whether it merely requires a local focus that limits the users' attention (e.g., when watching an engaging movie on a virtual handheld screen). The question then follows: What happens when this focus of attention or task space is indeed further away? Are stimuli prioritized when they are near the focus of attention (i.e., near the location where the motor task is occurring, despite being located further away), instead of prioritization of stimuli near the body?

In the present study we investigated the impact of user representations, i.e., the combination of the virtual visual representation of the user (i.e., virtual hand, virtual controllers, or no representation) and the input devices (i.e., Leap Motion, controllers, or keyboard). Future research should further investigate whether the observed effects are impacted when using different types of input devices or control mappings for the virtual representations. Moreover, in the present study we included a small sample size, since we opted for a within-groups experiment design. It is important for future research to investigate whether the present findings can be replicated including larger sample sizes. Future work on information presentation in VR should also carefully explore the use of other characteristics already applied for designing notifications in 2D (movement, size, color, duration), as well as VR-specific characteristics (semantics of specific locations in a room). Furthermore, in VR the user's attention may perhaps be captured through subtle ambient changes, e.g., in lighting or soundscape.

## 7 CONCLUSION

In this paper we explore information processing in a dual-task paradigm within an immersive virtual environment. Our results indicate that while engaged in a motor task, performance in detecting visual stimuli is higher if the visual information is presented near the user's body. However, a different processing of near and far away visual stimuli does not occur when the user is not concurrently performing a motor task and only focuses on detecting the visual information. Furthermore, we observed no difference in information processing when varying the type of user representations, with all user representations leading to similar processing of visual stimuli in a 3D immersive environment. The user representation did however impact motor performance, as well as the sense of embodiment, which are increased when having a body or tool representation.

### ACKNOWLEDGMENTS

This research has received funding from the European Union's Horizon 2020 research and innovation program under grant agreement #737087 (Levitate). The work was also partially supported by the Innovation Fund Denmark (MADE Digital project, IFD grant no. 6151-00006B). We thank Jan Milosch and Timm Seltmann for their help in conducting the study.

**Sofia Seinfeld.** Sofia Seinfeld is currently a lecturer in the Image Processing and Multimedia Technology Center from the Polytechnic University of Barcelona. Previously she worked as a postdoctoral researcher at the University of Bayreuth. She has a bachelor's degree in Psychology (2011) and master's degree in Clinical and Health Psychology (2015) from the University of Barcelona (2011). Sofia earned a PhD in Virtual Reality and Psychology (2017) from the University of Barcelona, working as a researcher in the Event Lab and Institut d'Investigacions Biomèdiques August Pi i Sunyer (IDIBAPS). As part of her research, she has published papers on topics related to the use of immersive virtual reality and embodiment to tackle problems such as domestic violence, racism, and fear of heights. Currently, her research focuses on exploiting multisensory integration principles and feedback to enhance human-computer interaction in virtual reality and levitation-based interfaces.

**Tiare Feuchtner.** Tiare Feuchtner is currently a postdoctoral researcher at Aarhus University and TU Wien. She earned her bachelor's degree in Media Informatics from TU Wien (2011), her master's in Computer Science from TU Berlin (2015) and was awarded a PhD in Computer Science from Aarhus University (2018). In her early career she was junior researcher at the Austrian Institute of Technology and the Telekom Innovation Laboratories in Berlin. Tiare has published several papers on various topics, including public display interaction, and embodiment in immersive environments (VR). Her current research interests include digital assistance approaches for the manufacturing industry through virtual and augmented reality, as well as embodied user interfaces for interaction in immersive virtual environments.

**Johannes Pinzek.** Johannes Pinzek is currently studying for a bachelor's degree in Biology and Computer Science in the University of Bayreuth.

**Jörg Müller.** Jörg Müller is a professor of Computer Science at the University of Bayreuth, Germany. He holds a bachelor's (2003) and a master's degree (2005) in Computer Science from Saarland University, Germany. He earned his PhD in Human-Computer Interaction (2009) from the University of Münster, Germany. He has worked as a Senior Researcher at Deutsche Telekom Laboratories, Berlin, as a Guest Professor at Berlin University of the Arts, and as an Associate Professor at Aarhus University in Denmark, before taking his current position at the University of Bayreuth. He has published more than 100 papers in the area of human-computer interaction. His current research is the understanding of movement in human-computer interaction.